\documentclass[]{aastex63}

\usepackage{float}
\usepackage{multirow}
\received{\today}
\revised{\today}
\accepted{\today}
\submitjournal{ApJ}

\shorttitle{Asymmetries of Sunspot Groups During their Decay}
\shortauthors{J. Murak\"ozy}

\begin{document}
	
	\title{Variations of the Internal Asymmetries of Sunspot Groups During their Decay}
	
	\correspondingauthor{Judit Murak\"ozy}
	\email{murakozy.judit@epss.hu}
	
	\author[0000-0001-6920-259X]{Judit Murak\"ozy}
	\affiliation{Institute of Earth Physics and Space Science (ELKH EPSS)\\
		Sopron, Hungary}
	
	\affiliation{Konkoly Observatory\\ Research Centre for Astronomy and Earth Sciences (ELKH)\\
		Konkoly-Thege Mikl\'os \'ut 15-17.\\
		Budapest, 1121, Hungary}
	
	
	\begin{abstract}
		The aim of the present study is to show the varying asymmetries during the decay of sunspot groups. The source of input data is the SoHO/MDI–Debrecen Database (SDD) sunspot catalog that contains the magnetic polarity data for time interval 1996-2010. Several types of asymmetries were examined on the selected sample of 142 sunspot groups. The leading-following asymmetry increases in three phases during the decay and exhibits anticorrelation with size. It is also related to a hemispheric asymmetry, during the decay the area asymmetry index has higher values in the southern hemisphere which may be due to the higher activity level in the southern hemisphere in cycle 23. The total umbral area is inversely proportional to the umbra/penumbra ratio but it is directly proportional to the umbral decay rate. During the decay the umbra/penumbra (U/P) ratio decreases unambiguously in the trailing parts but in most cases in the leading parts as well. The U/P variation is a consequence of the different depths of umbral and penumbral fields.
	\end{abstract}
	\keywords{Sunspots, Sun: activity --- Sun: magnetic field}

\section{Introduction}
           \label{S-Introduction}

The development of solar active regions until the maximum state and their decay afterwards are governed by different mechanisms. The detailed description of these events is important for the understanding of the complex system of interactions between solar magnetic and velocity fields. The investigation of sunspot decay started with the study of single spots. \citet{1946MNRAS.106..218C} published the theoretical consideration that the decay of a spot cannot be caused merely by diffusion, the surrounding velocity fields have to play a definitive role in it. This has been worked out in detail in the paper by \citet{1997SoPh..176..249P}. \citet{1946MNRAS.106..218C} also compared the evolutionary curve of two groups with their magnetic-life histories and obtained that the magnetic field increases very rapidly and reaches its maximum almost in the same time as the area. \citet{2005ApJ...635..659H} presented details of the moving magnetic features (MMFs) by tracking eight sunspots. MMFs are small knots around the sunspots with roughly 10$^3$ km diameter which transport magnetic flux away from the spots. \citet{2009SoPh..258...13K} studied the convective motions and the sunspot decay on a sample of 788 active regions and found that the strong upflow changes to downflow at a certain depths during the decay. \citet{2005ApJ...635..659H} pointed out that there are more MMFs around the larger spots by studying eight sunspots.

The decay of sunspot groups is a more complex series of events and interactions. 
In the model of \citet{1975Ap&SS..34..347P} the sunspot groups start to decay when the flux ropes lose their twists. Thus unwinding of the flux ropes frays the rope itself. The large, long-lived sunspots are bound with an annular moat and measured an outward velocity in the moat. When the spots start to decay small magnetic knots can be observed moving outward across the moat. These knots carry away the flux from the spot. This model describes strong plasma control of the flux tube. \citet{2017ApJ...842....3N} studied the decay in some cases and found higher decay rate in the following part and obtained a relation between the rate of the MMFs and the leading/following decay rates. The average value of the decaying flux is in agreement with the rate obtained by \citet{2005ApJ...635..659H}. \citet{2009SoPh..258...13K} studied the convective motions and the sunspot decay on a sample of 788 active regions and found that the strong upflow changes to downflow at a certain depths during the decay. \citet{2007ApJ...671.1013D} observed a decaying follower sunspot over six days and pointed out that the umbra/total areas increased from 15.9 \% to 19 \% during the decay, showing that the umbra decays slower than the penumbra. Although the decay rate was found as almost constant, the decay process is not uniform. They described the decay as a three steps process. Firstly, the fragmentation of the sunspot can be observed, then the flux cancellation of MMFs that encounter the opposite polarity network at the edge of the moat region, while at the end the flux is transported by MMFs. \citet{2014SoPh..289.1531G} followed the evolution of four ARs by using intensitygrams obtained by the Helioseismic and Magnetic Imager (HMI) onboard Solar Dynamics Observatory (SDO). They found that the largest contributor to the total area decay rates of spots is the decay rates of penumbra, while the umbral decay rates are much lower. These results are in agreement with the observations of \citet{2010AN....331..563S}.

\citet{2014ApJ...785...90R} in their theoretical simulation obtained that lifetimes of sunspots are too short in the absence of a penumbra. They concluded that the penumbra may either stabilize the surface layers of sunspots or it delays the fragmentation of the subsurface layers of sunspots.

There are some works that analyze different asymmetries during the evolution of active regions. \citet{2012Ap&SS.338..217J} examined the rates of growth and decay of sunspot groups and pointed out differences between the hemispheric values. Some other works studied the tilt angles of sunspot groups and their variations during the decay and observed a difference between the northern and southern values \citep{2012ApJ...758..115L, 2013SoPh..287..215M}, while \citet{1993SoPh..145...95H} investigated the dependence of the decay of sunspot groups on axial tilt angles. \citet{2014SoPh..289..563M} studied the area and number asymmetries of sunspot groups at their maximum states and pointed out an asymmetry in the compactness of groups, i.e., the number of spots tends to be smaller, while their mean area is larger in the leading part at the maximum phase.

\citet{1993A&A...274..521M} obtained that the total area to umbra area ratio is about 4–6, and noted that this parameter is independent of the evolutionary phase of the spot except the very last stage when the leading umbra is the only remnant of the sunspot. \citet{2007ApJ...671.1013D} also analyzed the UP/U ratio during the decay of the NOAA 10773 AR and found it varies between 5–6. \citet{2018ApJ...865...88C} studied the decay and the U/P ratio of sunspots during the Maunder Minimum and revealed that the value of the U/P ratio varies between 0.15–0.25, and the higher the U/P ratio the faster the decay of the sunspot.
\citet{2019SoPh..294...72J} obtained 5.5–6 for the P/U value and pointed out that this ratio is independent of cycle strength, latitude, and cycle phase. The results of \citet{2013SoPh..286..347H} are in agreement with this, however he found that this ratio increases with the increasing total sunspot group area. \citet{1990SoPh..129..191B} studied 126 sunspots observed around 1980 and pointed out the U/P value is 0.24 for small and 0.32 for large spots. \citet{2018SoPh..293..104C} found different behavior in the variation of U/P ratio of the small and large groups by using the Royal Greenwich Observatory (RGO) series. The larger groups do not show significant changes from year to year while the smaller groups do. \citet{1997rscc.book.....H} noted that the rate of sunspot decay is proportional to the convective velocity, which means that the higher the convective velocities the higher the U/P values and the faster the sunspot decay.

The aforementioned investigations dealt with the decay process of sunspot groups and some of them focused on the internal process as well. After the calculation of the distinct decay rates \citep{2021ApJ...908..133M} this study aims to describe the variation of the asymmetries within the sunspot groups during their decay.

\section{Data and methods}
\subsection{Observational data}
The present study has been made by using the SoHO/MDI--Debrecen Database (SDD) \citep{2016SoPh..291.3081B} which is one of the sunspot catalogs made in the Debrecen Observatory and contains sunspot data from 1996 until 2010. This database has about 1.5 hours temporal resolution which is allowed by the observations of the Solar and Heliospheric Observatory (SoHO) spacecraft and besides the area and position data contains also magnetic data for each observable sunspot group as well as for each sunspot within them. Thus, the leading and following parts of the sunspot groups can be distinguished, and the temporal resolution makes it possible to track the evolution of the groups and their parts. 

In order to select from this huge database those sunspot groups that are definitely in the decay phase, the following strict criteria were set. The sunspot groups should have two opposite polarity parts at the time of the maximum area. The development area has to be observed at least for two days while the decaying area has to be tracked at least during four consecutive days after the maximum and the first and last observed areas could be at most 40 \% of the maximum area. All the three areas (total area, leading and following areas) have to decrease during the decay phase, and these criteria were visually inspected in each case as well. The total number of the selected sample is 142.

\begin{figure}[H]
	\includegraphics[width=\textwidth]{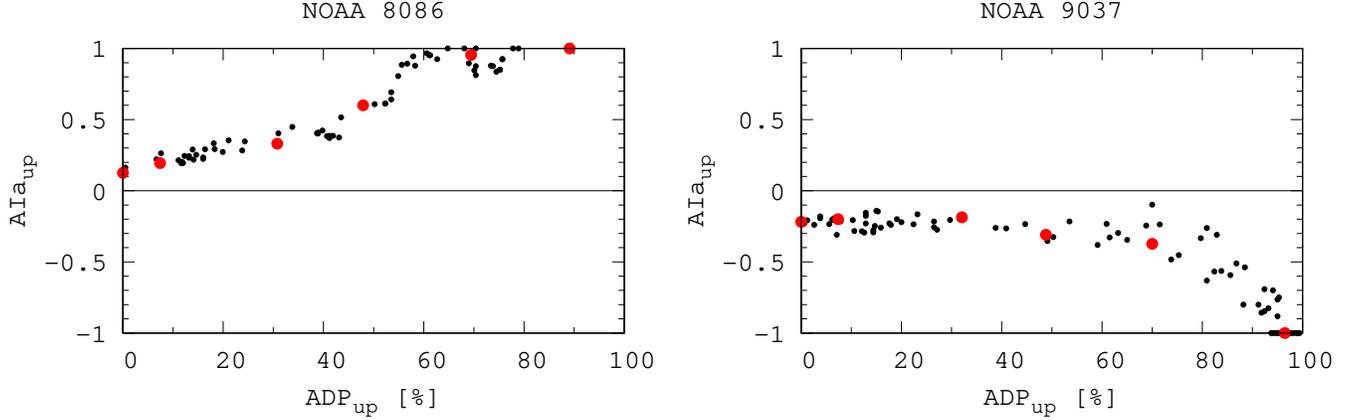}	
	\caption{Variation of the area asymmetry indices of AR 8086 (on the left) and 9037 (on the right) after their maximum areas during the decay phase. The red dots mark those data which are in Table~\ref{table:noaas}. }
	\label{fig:samplears}
\end{figure}

\begin{table}[H]
	\centering
	\begin{tabular}{ |p{1.0cm}|p{1.0cm}|p{1.0cm}|p{1.0cm}|p{1.0cm}||p{1.0cm}|p{1.0cm}|p{1.0cm}|p{1.0cm}|p{1.0cm}|  }
		\hline
		\multicolumn{1}{|c}{}&\multicolumn{3}{c}{NOAA 8086;  A$_{up}$=432 MSH}& \multicolumn{1}{c||}{}&\multicolumn{1}{c}{}&\multicolumn{3}{c}{NOAA 9037; A$_{up}$=340 MSH}& \multicolumn{1}{c|}{}\\
		\hline
		\hline
		$a_{up}$ & ADP$_{up} $ & lead. $a_{up}$ & foll. $a_{up}$ & $AIa_{up}$ & $a_{up}$ & ADP$_{up}$ & lead. $a_{up}$ & foll. $a_{up}$ & $AIa_{up}$ \\
		\hline
		432 & 0 & 243 & 189 & 0.125 & 340 & 0 & 133 & 207 & -0.218\\
		400 & 7.407 & 239 & 161 & 0.195 & 315 & 7.353 & 126 & 189 & -0.200\\
		299 & 30.787 & 199 & 100 & 0.331 & 231 & 32.059 & 94 & 137 & -0.186\\
		225 & 47.917 & 180 & 45 & 0.6 & 174 & 48.824 & 60 & 114 & -0.310\\
		132 & 69.444 & 129 & 3 & 0.955 & 102 & 70 & 32 & 70 & -0.373\\
		47 & 89.120 & 47 & NO & 1 & 12 & 96.471 & NO & 12 & -1\\
		\hline
	\end{tabular}
	\caption{Detailed data of the ARs of Fig.~\ref{fig:samplears}, i.e. AR 8086 (first set of six columns) and AR 9037 (last set of six columns), respectively.  $A_{up}$ is the maximum area of the group in millionths of solar hemipshere (MSH) at ADP 0. Columns of the sets are as follows. $a_{up}$ means the observed umbra+penumbra area measured in MSH, ADP is calculated by using Eq.~\ref{eq:adp} measured in \%, areas of the leading and following parts in MSH, while the last column describes the area asymmetry index calculated by using Eq.~\ref{eq:asymmetry}. NO means that leading or following sunspot already can not be observed at all.}
	\label{table:noaas}
\end{table}

\subsection{Method}
In this study the normalized asymmetry index (AI) is used.
The asymmetry index between the leading and following part of sunspot groups is calculated as
\begin{equation}
AIx=\frac{x_{l}-x_{f}}{x_{l}+x_{f}}.
\label{eq:asymmetry}
\end{equation}

where $x_{l}$ and $x_{f}$ mean the parameter of the leading and following parts, respectively. This property may be the area of the sunspot groups (a$_{up}$) or that of the umbrae (a$_u$).
The decay phase of the sunspot groups will be characterized by the area decay phase (hereafter ADP) which is determined for each observational time of the groups
\begin{equation}
ADP=\Big(1-\frac{a}{A}\Big)*100
\label{eq:adp}
\end{equation}
 where $A$ is the maximum area of the group and $a$ is the observed area. If the value of the asymmetry index is 0 that means the parameters of the leading and following parts are equal, while 1 and -1 mean that the following or leading part is missing and the relevant parameter only refers to the existing part. ADP=0 marks the maximum value of the sunspot group's area (see Table~\ref{table:noaas}).

These asymmetry indices are calculated for each observational time or area decay phase. Then the obtained asymmetry indices are  averaged over 10 percent bins of the ADP. Although in typical cases the mean size of the follower spots is smaller than that of the leader ones (left panel of Fig.~\ref{fig:samplears}), there are exceptions where the mean size of following spots is larger than that of the leader ones (right panel of Fig.~\ref{fig:samplears}) these are characterized by negative asymmetry index. These data, i.e., the ADP, and the AIa$_{up}$ can be found in Table~\ref{table:noaas} for the ARs NOAA 8086 and NOAA 9037. The table contains only six raws for each active region after their maximum area of the whole group. ADP=0 denotes the maximum area of sunspot groups.

\section{Results and Discussion}
In a previous paper \citep{2021ApJ...908..133M} it has been shown that the leading and following parts of sunspot groups decay with different rates. After the determination of their decay rates, the dynamics of the variation of the two parts are calculated. First of all, the area asymmetry index is studied for both the whole sunspot groups and only the umbrae.

\begin{figure}[H]
	\includegraphics[scale=0.9, angle=0]{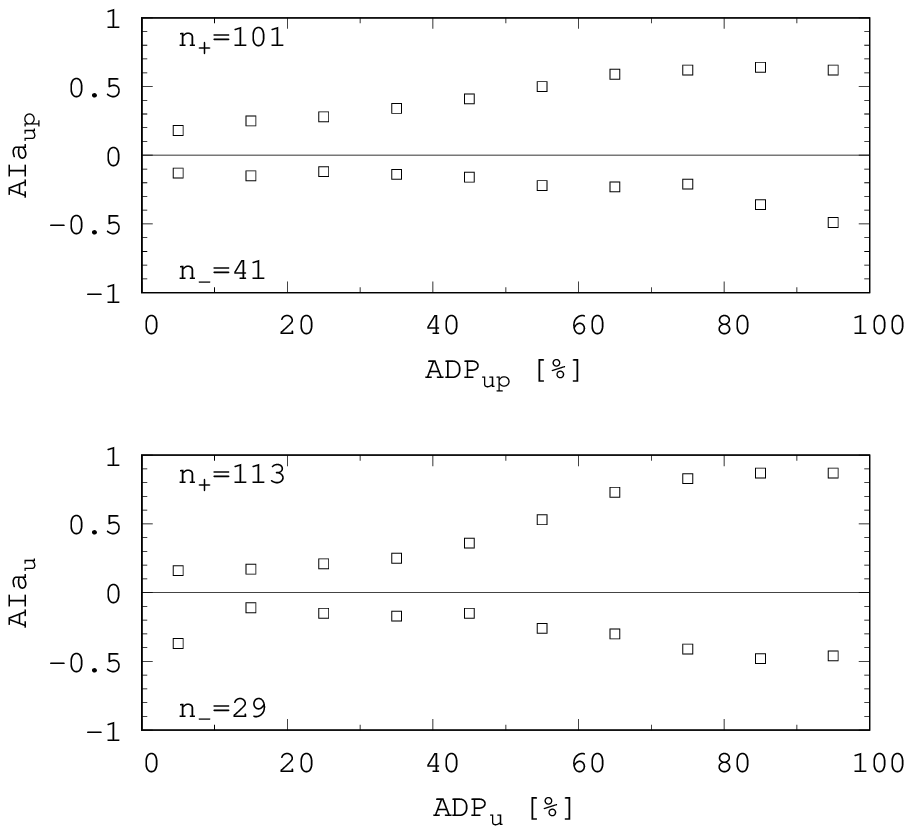}
	\includegraphics[scale=0.9, angle=0]{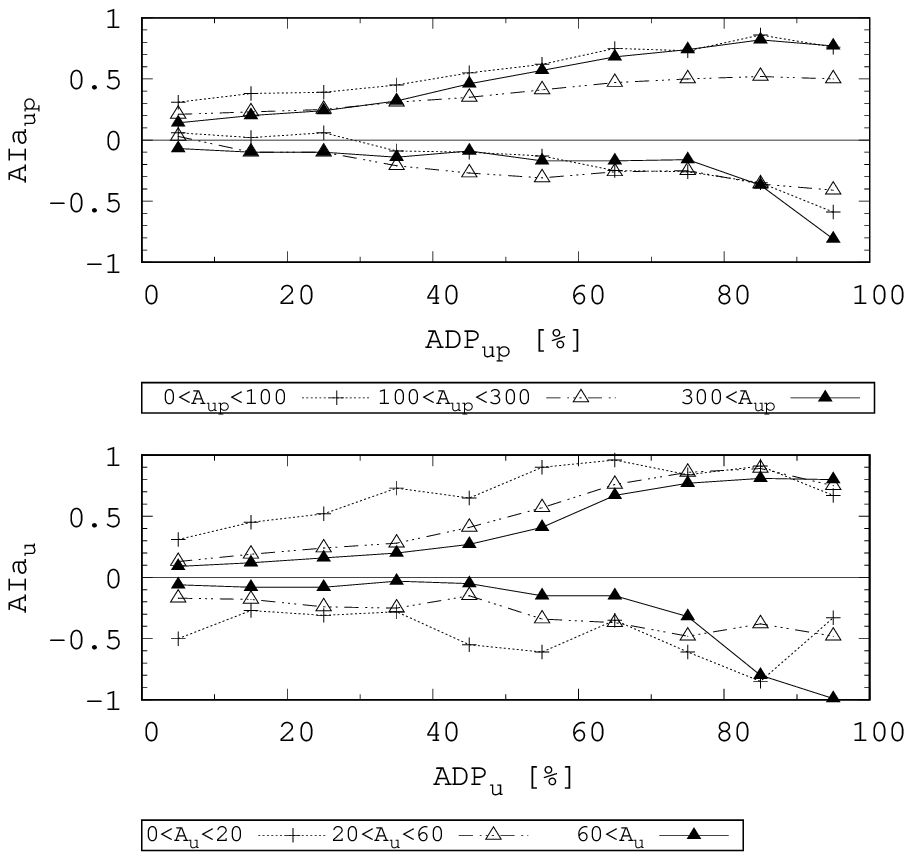}
	\caption{Left: Dependence of the normalized asymmetry index on the area decay phase calculated for the total (umbra+penumbra) areas (top panel) and the umbral area (bottom panel). The numbers of the positive/negative cases are marked at the left corners of the panels, the numbers of positive cases are significantly higher. Right: The same as in the left panels but the asymmetry indices are calculated for three area ranges. \label{fig:aiacucup}}
\end{figure}

The panels of Fig.~\ref{fig:aiacucup} show the combined history of the decay and the asymmetry variation of sunspot groups after their maximum state, the area asymmetry indices are averaged over 10 percent bins of the area decay phase. The sunspot groups of positive and negative asymmetries (according to Eq.~\ref{eq:asymmetry}) are plotted separately. The left columns show all sunspot group sizes together. It is a common property of the diagrams that the increase of the asymmetry begins at about 35\% of the area decay but it is conspicuous that after this the variation is steepest for the umbral areas of sunspot groups of positive asymmetry (upper half of the lower frame in the left column). This can be considered to be the most typical case, the process starts at the maximum state with very small asymmetry that reaches the value of +1 at the end with a single leading spot. In contrast, the sunspot groups of negative area asymmetry (lower frame, lower diagram) do not end with asymmetry of -1, i.e. with only a follower polarity. 

Overall, it can be said that the leading/following area ratio of the maximum which is near by 0, means the area of both parts is almost equal and is preserved during the first phase of the decay. This is followed by the steeper variation when the smaller part of the group starts to disappear and in the last phase of the decay it almost or totally disappears and the total umbral area is dominated by only the part with a larger area.
During the first phase, the area of the leading/following part is about 50 percent larger than the area of the following/leading part. This ratio increases during the decomposition and reaches about the 0.7 in the case of the total area, while this value is higher, about 0.9 in the case of the umbrae. In an earlier work \citep{2014SoPh..289..563M} similar variation has been observed in the asymmetric emergence of the leading-following parts.
\citet{2007ApJ...671.1013D} also described the decay as a three steps process but only on the decay of one following sunspot. They identified the three steps with fragmentation, the flux cancellation and the flux transport by MMFs, respectively.
\\The right column shows the same variations in three area ranges. For umbra+penumbra areas they are indicated in the upper panel, the umbral area ranges are defined as the one fifths of the total area ranges as in a previous paper \citep{2021ApJ...908..133M} this ratio was found at the time of maximum. It is conspicuous that the most typical decay pattern is exhibited by the largest sunspot groups where the asymmetry is close to zero at the maximum area, its absolute value starts rising around one third of the ADP of the group and at the end it reaches the final values of about +1 or -1. The time profiles of the smaller groups are more flattened, especially those of negative asymmetry. 

\begin{figure}
	\includegraphics[scale=1.00, angle=0]{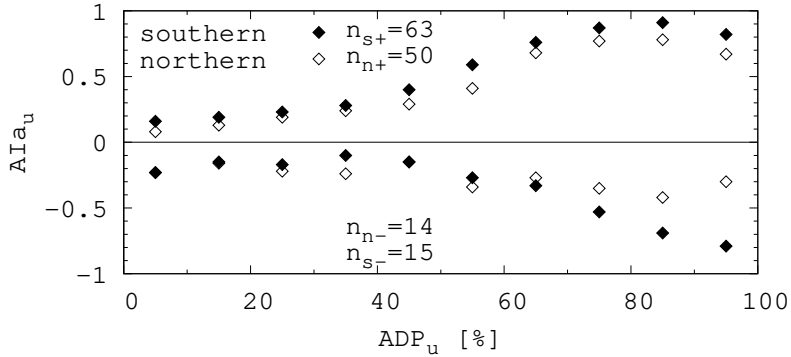}
	\caption{Hemispheric variation of the normalized umbral area asymmetry index calculated between the leading and following parts (AIa$_u$) during the decay of the sunspot groups. The northern hemispheric values are marked by full rhombuses, the southern values are depicted by empty rhombuses. The top and bottom panels depict the cases of positive/negative asymmetries, the two hemispheres are distinguished in both subsets. \label{fig:NS}}
\end{figure}

The two hemispheres have been examined separately. Fig.~\ref{fig:NS} shows the umbral ADP variation during the decay phase by distinguishing the types of asymmetries and the hemispheres, thus the diagram is a more detailed version of the lower left frame of Fig.~\ref{fig:aiacucup}. The additional information can be read from the lower frame of the diagram showing the data of groups of negative asymmetry. Here the data of the southern hemisphere follow the standard time profile: unambiguous strengthening of the asymmetry starting at one third of the decay time interval and ending close to -1, while the data set of the northern hemisphere exhibits a weaker variation. This explains the similarly weak variation of the combined North-South data in the negative domain of the lower left frame in Fig.~\ref{fig:aiacucup}.

These distinctions permit a conjecture. The parallel course of asymmetry variation and decay takes the above formulated standard form typically in large sunspot groups even in the less frequent cases of negative asymmetry. This can be seen in the lower right frame of Fig.~\ref{fig:aiacucup} in its positive and negative halves. On the other hand Fig.~\ref{fig:NS} also presents an indirect evidence for this, the time profile of negative asymmetries in the southern hemisphere corresponds to this pattern. The southern hemisphere is more active in most of the solar cycle 23 covered by the applied data \citep{2013ApJ...768..188C}, it also predominates in the applied sample shown in Fig.~\ref{fig:NS}, furthermore, the umbral area asymmetry index is always higher in  the southern hemisphere. This may imply that the strong flux ropes emerging from the strong toroidal magnetic fields, presumably from deeper layers are subjects to a different set of impacts than the smaller sunspot groups. This may be a consequence of the higher sunspot activity in cycle 23. Several hemispheric asymmetries have been pointed out e.g. in the decay rate \citep{2021ApJ...908..133M} or in the tilt angles of ARs \citep{2012ApJ...758..115L, 2013SoPh..287..215M}.

\begin{figure}
	\includegraphics[scale=1.0, angle=0]{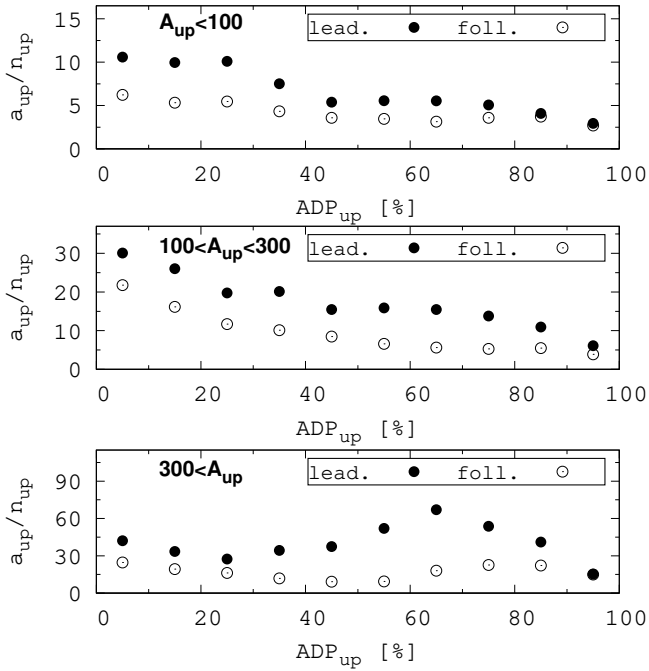}
	\caption{ Mean sunspot area of the leading part (dots) and following part (circle) averaged over 10 percent bins of the decay phase of the whole group. 0\% means the maximum area of the groups. The three panels show the variations of three different area ranges. \label{fig:avgspotsize}}
\end{figure}

The average area of the sunspot group is also studied (Fig.~\ref{fig:avgspotsize}). The three panels concern different sizes of sunspot groups from the smallest ones (A$_{up} <$100 MSH) to the biggest groups (A$_{up} >$300 MSH). The shapes of the courses are similar in the cases of the smallest and the medium size sunspot groups. These groups show that the average sizes of sunspots of the leading and following parts decrease simultaneously until 45 \% of the ADP$_{up}$ where reach they their plateaus and after 75 \% of the ADP$_{up}$ they decrease again. The average sizes of the leading spots are higher during the whole decay in all three cases.
The course of the decay is somewhat different in the case of the biggest groups. Here the average sizes of spots show an increasing trend after the short first phase of decay. This is caused by the sudden drop in the number of spots, i.e. the smallest spots disappear around 70\% of the ADP$_{up}$, while the largest spots survive. This is more pronounced in the case of the leading spots. At the end of the ADP$_{up}$ the average sizes of sunspots will be nearly the same in each case.
The area ratio of the leading/following spots is different in these three area ranges. The ratio decreases toward the end of decay in the cases of the smallest and the middle groups, but in the case of the biggest groups this ratio increases in the middle of ADP. 
It can also be seen that the higher the area of groups the smaller the ratio between the leading and following spots around the maxima. The total area of leading spots is about twice larger than that of the following ones in smallest groups and this ratio is about 1.5 in medium and large groups.

The penumbrae (P) are formed by the strongly inclined field lines of sunspots at the surface layers \citep{2021ApJ...907..102P} so their decay may be controlled by processes different from those of the umbrae (U) whereby influencing the variation of their area ratio.

\begin{figure}
	\includegraphics[scale=0.9, angle=0]{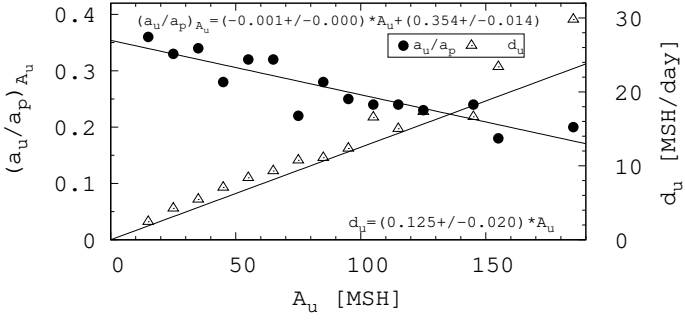}
	\caption{The a$_u$/a$_{p}$ value at the time of maximum umbrae averaged over 10 MSH bins as a function of maximum umbral area depicted with black dots. The umbral decay rate \citep{2021ApJ...908..133M} averaged also over 10 MSH bins as a function of maximum umbral area plotted with empty triangles. 
		\label{fig:uperp}}
\end{figure} 
\noindent
Fig.~\ref{fig:uperp} shows the U/P area ratios of spots in the states of maximum umbral areas plotted with black dots. Their distribution exhibits a clear inverse relationship with the maximum area, larger spots have relatively smaller umbrae with respect to the penumbrae. The other diagram, the  decay rates vs. maximum area is taken from an earlier paper \citep{2021ApJ...908..133M}, the data are plotted with empty triangles. Both datasets are averaged over 10 MSH bins of maximum umbral area. The opposite trends of the two diagrams is conspicuous, larger umbrae decay faster than smaller ones and their areas with respect to their penumbrae are smaller than in small sunspots. This is in agreement with the theoretical result of \citet{2014ApJ...785...90R} that the larger penumbra stabilizes the sunspot; but it contradicts to the results of \citet{1990SoPh..129..191B} and \citet{2018SoPh..293..104C} and \citet{2013SoPh..286..347H}. This dependence is also in contrast to \citet{1997rscc.book.....H}, who found a linear relationship between the U/P values and the decay rates.

\begin{figure}
	\includegraphics[scale=0.8, angle=0]{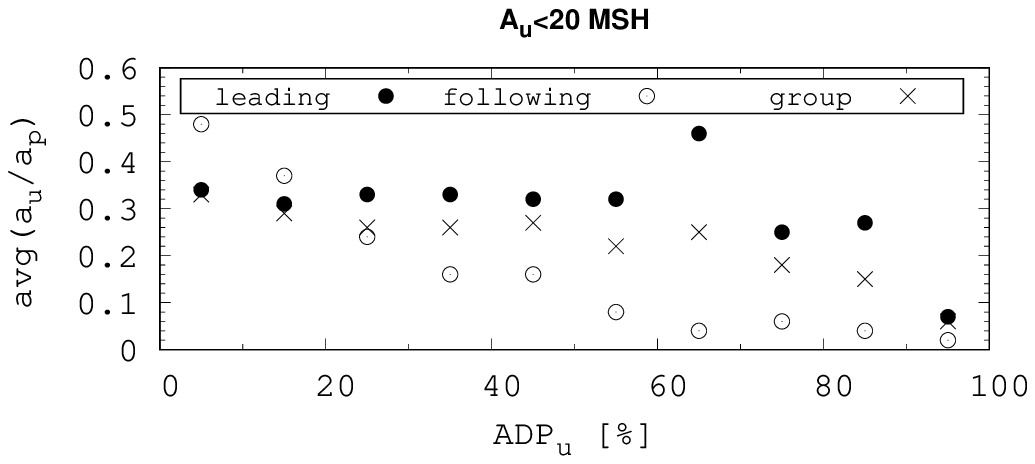}
	\includegraphics[scale=0.8, angle=0]{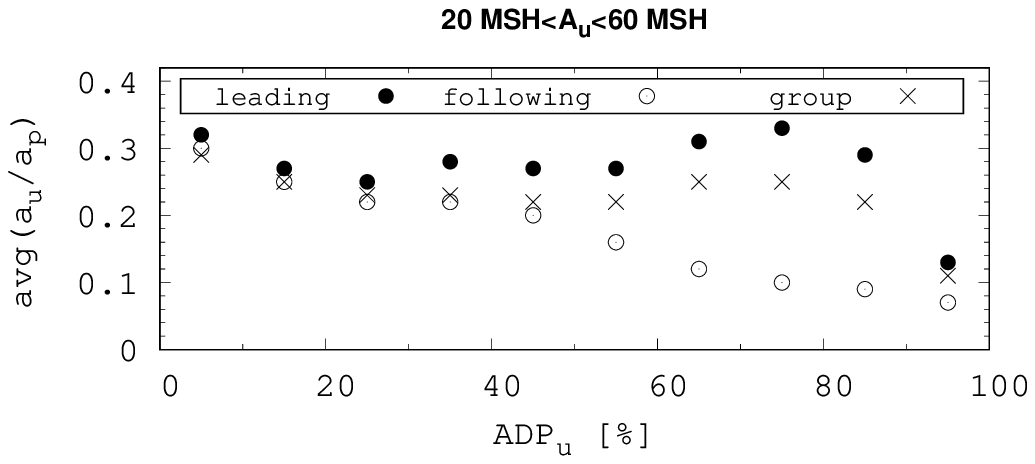}
	\includegraphics[scale=0.8, angle=0]{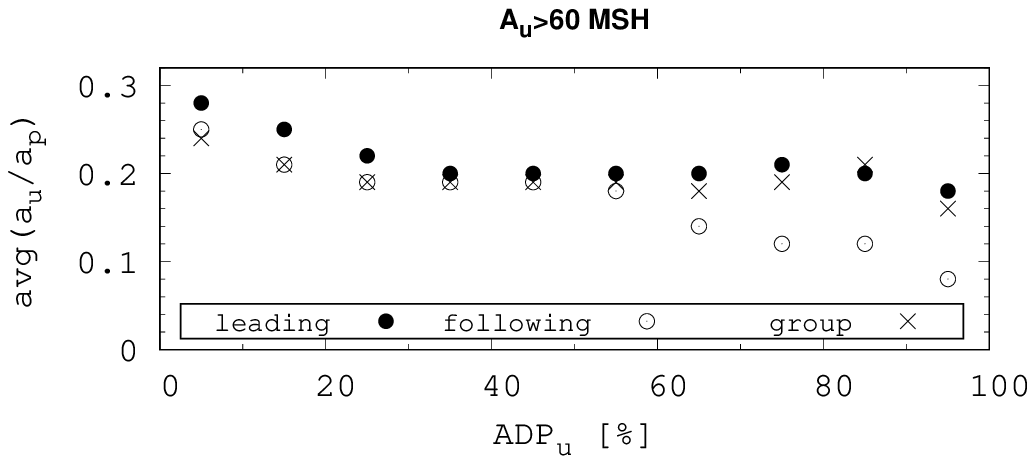}
	\caption{Umbra and penumbra ratios of the whole groups (crosses) and their leading (dots) and following (empty circles) parts as a function of the umbral area decay phase. The values of a$_u$/a$_{p}$ are averaged over 10 percent bins of the umbral ADP. 0\% marks the maximum umbral area. Left top panel: A$_u\ge$ 20 MSH, right top panel: 20 MSH $\le$ A$_u\le$ 60 MSH, left bottom panel: A$_u\ge$ 60 MSH.
	\label{fig:uperpdepa}}
\end{figure}

\noindent
The decay process also exerts an impact on the variation of U/P ratio as is shown in Fig.~\ref{fig:uperpdepa}. The sample is divided into three groups of sizes as in Fig.~\ref {fig:avgspotsize}, the data of leading and following parts as well as the entire groups are plotted separately and the decay is represented again in a standard time interval normalized to the lengths of decays starting with the maximum area (at zero Area Decay Phase, ADP). The most striking feature of these diagrams is the definite decrease of the U/P ratio during the decay in the following parts of the sunspot groups (indicated with empty circles), which means that in the trailing regions the umbrae disappear more quickly than the penumbrae. In the leading parts the decreasing trend is also present but with some temporary strengthening. This may be due to the typically larger leading umbrae which may be more resistant to the disintegrating due to external impacts than those of the trailing part. Anyhow, the overall trend is that the deeper rooted umbrae are more intensively exposed to the decomposing impact of the external processes than the penumbrae close to the surface layers.

The courses of the decays of the leading part and the whole group are similar in each case. As a result of this U/P study one can conclude that the smaller the sunspot group area the higher the U/P ratio and the difference between values of the leading and following parts. Moreover the smaller the sunspot groups the higher the variation of this ratio during the decay.

\section{Summary and conclusion}
During the decay process of sunspot groups several characteristic variations happen in their internal structures.
The results can be summarized as follows.

(i) The sunspot group’s decay can be divided into three parts where the leading/following asymmetries vary with different rates (left panel of Fig.~\ref{fig:aiacucup}). This asymmetry is almost constant in the first phase of the decay, its ratio slightly varies and is preserved from the time of the maximum of the groups. Then that varies faster during the middle phase of the decay. After this steeper variation the area asymmetry seems to be stabilized.

(ii) The variation of the leading-following umbral area asymmetry depends on the sunspot group’s maximum size. It rises earlier in small groups which contain typically small spots that disappear more quickly. The asymmetry variation of the total (U+P) area is less sensitive to the disintegrating impacts, the variation of curves of the penumbrae are more flattened than that of the umbrae (right panel of Fig.~\ref{fig:aiacucup}). 

(iii) The leading/following area asymmetry also exhibits hemispheric difference (Fig.~\ref{fig:NS}). During the sunspot group’s decay the area asymmetry index has higher values in the southern hemisphere.

(iv) The umbra–penumbra ratio at the time of maximum umbra exhibits anticorrelation with the area (Fig.~\ref{fig:uperp}). The decay rate and the U/P ratio are inversely proportional.

(v) The variation of the umbra-penumbra ratio during the decay depends on the maximum area of the group and also on the leading or following positions of the spots (Fig.~\ref{fig:uperpdepa}). The variation is typically a decrease which is the strongest in the following parts of small groups, but in the leading parts some temporary strengthenings may occur. The presented processes imply that the umbrae are more exposed to disintegrating effects than the penumbrae which are only affected by surface velocity fields.

The behavior of the larger groups differs from that of medium and small groups in many ways e.g. the average area of sunspots within the groups, the U/P ratio, as well as the variation of the area asymmetry index. The physical conditions affecting them are different. The leading--following area asymmetry changes rapidly, the following spots vanish earlier but the ratio between the umbra--penumbra hardly changes mainly in the case of the leading spots. This means that the decay of the larger groups is a smooth process, while the small groups behave more chaotically.
\section*{Acknowledgements}

This research has received funding from National Research, Development and Innovation Office -- NKFIH, 129137. Thanks are due to Dr. Andr\'as Ludm\'any for reading and discussing the manuscript and the anonymous referee whose comments made this article easier to understand.


\bibliography{Murakozy_2020_DECAY3_paper_accepted_in_ApJ_arxiv}{}

\begin{thebibliography}{}
\expandafter\ifx\csname natexlab\endcsname\relax\def\natexlab#1{#1}\fi
\providecommand{\url}[1]{\href{#1}{#1}}
\providecommand{\dodoi}[1]{doi:~\href{http://doi.org/#1}{\nolinkurl{#1}}}
\providecommand{\doeprint}[1]{\href{http://ascl.net/#1}{\nolinkurl{http://ascl.net/#1}}}
\providecommand{\doarXiv}[1]{\href{https://arxiv.org/abs/#1}{\nolinkurl{https://arxiv.org/abs/#1}}}

\bibitem[{{Baranyi} {et~al.}(2016){Baranyi}, {Gy{\H{o}}ri}, \&
  {Ludm{\'a}ny}}]{2016SoPh..291.3081B}
{Baranyi}, T., {Gy{\H{o}}ri}, L., \& {Ludm{\'a}ny}, A. 2016, \solphys, 291,
  3081, \dodoi{10.1007/s11207-016-0930-1}

\bibitem[{{Brandt} {et~al.}(1990){Brandt}, {Schmidt}, \&
  {Steinegger}}]{1990SoPh..129..191B}
{Brandt}, P.~N., {Schmidt}, W., \& {Steinegger}, M. 1990, \solphys, 129, 191,
  \dodoi{10.1007/BF00154373}

\bibitem[{{Carrasco} {et~al.}(2018{\natexlab{a}}){Carrasco},
  {Garc{\'\i}a-Romero}, {Vaquero}, {Rodr{\'\i}guez}, {Foukal}, {Gallego}, \&
  {Lef{\`e}vre}}]{2018ApJ...865...88C}
{Carrasco}, V.~M.~S., {Garc{\'\i}a-Romero}, J.~M., {Vaquero}, J.~M., {et~al.}
  2018{\natexlab{a}}, \apj, 865, 88, \dodoi{10.3847/1538-4357/aad9f6}

\bibitem[{{Carrasco} {et~al.}(2018{\natexlab{b}}){Carrasco}, {Vaquero},
  {Trigo}, \& {Gallego}}]{2018SoPh..293..104C}
{Carrasco}, V.~M.~S., {Vaquero}, J.~M., {Trigo}, R.~M., \& {Gallego}, M.~C.
  2018{\natexlab{b}}, \solphys, 293, 104, \dodoi{10.1007/s11207-018-1328-z}

\bibitem[{{Chowdhury} {et~al.}(2013){Chowdhury}, {Choudhary}, \&
  {Gosain}}]{2013ApJ...768..188C}
{Chowdhury}, P., {Choudhary}, D.~P., \& {Gosain}, S. 2013, \apj, 768, 188,
  \dodoi{10.1088/0004-637X/768/2/188}

\bibitem[{{Cowling}(1946)}]{1946MNRAS.106..218C}
{Cowling}, T.~G. 1946, \mnras, 106, 218, \dodoi{10.1093/mnras/106.3.218}

\bibitem[{{Deng} {et~al.}(2007){Deng}, {Choudhary}, {Tritschler}, {Denker},
  {Liu}, \& {Wang}}]{2007ApJ...671.1013D}
{Deng}, N., {Choudhary}, D.~P., {Tritschler}, A., {et~al.} 2007, \apj, 671,
  1013, \dodoi{10.1086/523102}

\bibitem[{{Gafeira} {et~al.}(2014){Gafeira}, {Fonte}, {Pais}, \&
  {Fernandes}}]{2014SoPh..289.1531G}
{Gafeira}, R., {Fonte}, C.~C., {Pais}, M.~A., \& {Fernandes}, J. 2014,
  \solphys, 289, 1531, \dodoi{10.1007/s11207-013-0440-3}

\bibitem[{{Hagenaar} \& {Shine}(2005)}]{2005ApJ...635..659H}
{Hagenaar}, H.~J., \& {Shine}, R.~A. 2005, \apj, 635, 659,
  \dodoi{10.1086/497367}

\bibitem[{{Hathaway}(2013)}]{2013SoPh..286..347H}
{Hathaway}, D.~H. 2013, \solphys, 286, 347, \dodoi{10.1007/s11207-013-0291-y}

\bibitem[{{Howard}(1993)}]{1993SoPh..145...95H}
{Howard}, R.~F. 1993, \solphys, 145, 95, \dodoi{10.1007/BF00627985}

\bibitem[{{Hoyt} \& {Schatten}(1997)}]{1997rscc.book.....H}
{Hoyt}, D.~V., \& {Schatten}, K.~H. 1997, {The role of the sun in climate
  change}

\bibitem[{{Javaraiah}(2012)}]{2012Ap&SS.338..217J}
{Javaraiah}, J. 2012, \apss, 338, 217, \dodoi{10.1007/s10509-011-0932-2}

\bibitem[{{Jha} {et~al.}(2019){Jha}, {Mandal}, \&
  {Banerjee}}]{2019SoPh..294...72J}
{Jha}, B.~K., {Mandal}, S., \& {Banerjee}, D. 2019, \solphys, 294, 72,
  \dodoi{10.1007/s11207-019-1462-2}

\bibitem[{{Komm} {et~al.}(2009){Komm}, {Howe}, \& {Hill}}]{2009SoPh..258...13K}
{Komm}, R., {Howe}, R., \& {Hill}, F. 2009, \solphys, 258, 13,
  \dodoi{10.1007/s11207-009-9398-6}

\bibitem[{{Li} \& {Ulrich}(2012)}]{2012ApJ...758..115L}
{Li}, J., \& {Ulrich}, R.~K. 2012, \apj, 758, 115,
  \dodoi{10.1088/0004-637X/758/2/115}

\bibitem[{{Martinez Pillet} {et~al.}(1993){Martinez Pillet}, {Moreno-Insertis},
  \& {Vazquez}}]{1993A&A...274..521M}
{Martinez Pillet}, V., {Moreno-Insertis}, F., \& {Vazquez}, M. 1993, \aap, 274,
  521

\bibitem[{{McClintock} \& {Norton}(2013)}]{2013SoPh..287..215M}
{McClintock}, B.~H., \& {Norton}, A.~A. 2013, \solphys, 287, 215,
  \dodoi{10.1007/s11207-013-0338-0}

\bibitem[{{Murak{\"o}zy}(2021)}]{2021ApJ...908..133M}
{Murak{\"o}zy}, J. 2021, \apj, 908, 133, \dodoi{10.3847/1538-4357/abcfba}

\bibitem[{{Murak{\"o}zy} {et~al.}(2014){Murak{\"o}zy}, {Baranyi}, \&
  {Ludm{\'a}ny}}]{2014SoPh..289..563M}
{Murak{\"o}zy}, J., {Baranyi}, T., \& {Ludm{\'a}ny}, A. 2014, \solphys, 289,
  563, \dodoi{10.1007/s11207-013-0416-3}

\bibitem[{{Norton} {et~al.}(2017){Norton}, {Jones}, {Linton}, \&
  {Leake}}]{2017ApJ...842....3N}
{Norton}, A.~A., {Jones}, E.~H., {Linton}, M.~G., \& {Leake}, J.~E. 2017, \apj,
  842, 3, \dodoi{10.3847/1538-4357/aa7052}

\bibitem[{{Panja} {et~al.}(2021){Panja}, {Cameron}, \&
  {Solanki}}]{2021ApJ...907..102P}
{Panja}, M., {Cameron}, R.~H., \& {Solanki}, S.~K. 2021, \apj, 907, 102,
  \dodoi{10.3847/1538-4357/abccbf}

\bibitem[{{Petrovay} \& {van Driel-Gesztelyi}(1997)}]{1997SoPh..176..249P}
{Petrovay}, K., \& {van Driel-Gesztelyi}, L. 1997, \solphys, 176, 249,
  \dodoi{10.1023/A:1004988123265}

\bibitem[{{Piddington}(1975)}]{1975Ap&SS..34..347P}
{Piddington}, J.~H. 1975, \apss, 34, 347, \dodoi{10.1007/BF00644803}

\bibitem[{{Rempel} \& {Cheung}(2014)}]{2014ApJ...785...90R}
{Rempel}, M., \& {Cheung}, M.~C.~M. 2014, \apj, 785, 90,
  \dodoi{10.1088/0004-637X/785/2/90}

\bibitem[{{Schlichenmaier} {et~al.}(2010){Schlichenmaier}, {Bello
  Gonz{\'a}lez}, {Rezaei}, \& {Waldmann}}]{2010AN....331..563S}
{Schlichenmaier}, R., {Bello Gonz{\'a}lez}, N., {Rezaei}, R., \& {Waldmann},
  T.~A. 2010, Astronomische Nachrichten, 331, 563,
  \dodoi{10.1002/asna.201011372}

\end{thebibliography}
\bibliographystyle{aasjournal}

\end{document}